\documentstyle[11pt]{article}

\newcommand{\ccr}[2]{{[} {#1},{#2} {]} }        

\newcommand{\CMP}[1]{{\em Commun. Math. Phys.} {\bf {#1}}}

\newcommand{\NP}[1]{{\em Nucl.~Phys.~\bf B} {\bf {#1}}}

\newcommand{\PL}[1]{{\em Phys. Lett.} {\bf {#1}}}
\newcommand{\PR}[2]{{\em Phys. Rev.} {#1} {\bf {#2}}}

\newcommand{\eq}{\begin{equation}}
\newcommand{\eqend}{\end{equation}}
\newcommand{\eqa}{\begin{eqnarray}}
\newcommand{\neqa}{\begin{eqnarray*}}
\newcommand{\eqaend}{\end{eqnarray}}
\newcommand{\neqaend}{\end{eqnarray*}}
\newcommand{\nonu}{\nonumber \\ \nopagebreak}
\newcommand{\bma}[1]{\begin{array}{#1}}
\newcommand{\ema}{\end{array}}
\newcommand{\bc}{\begin{center}}
\newcommand{\ec}{\end{center}}

\newcommand{\GG}{\underline{G}}
\newcommand{\CC}{\underline{C}}
\newcommand{\ZZ}{\underline{Z}}
\newcommand{\PP}{\underline{P}}
\renewcommand{\gg}{\underline{g}}
\newcommand{\HH}{{\underline{H}}}
\newcommand{\hh}{{\underline{h}}}
\newcommand{\rtimes}{\mathbin{>\mkern-4mu<\mkern-7mu%
\hbox{\vrule width0.3pt height5.5pt depth0.9pt}}}
\newcommand{\U}{{\rm U}}
\newcommand{\SU}{{\rm SU}}

\newcommand{\Ref}[1]{(\ref{#1})}
\newcommand{\dd}{{\rm d}}
\newcommand{\ee}[1]{\mbox{{\rm e}}^{#1}}
\newcommand{\Exp}[1]{\exp{\left(#1\right)}}
\newcommand{\ii}{{\rm i}}
\newcommand{\R}{{\rm I\kern-.2emR}}

\newcommand{\Z}{{\sf Z} \! \! {\sf Z}}
\newcommand{\f}{\frac}
\newcommand{\cH}{{\cal H}}
\newcommand{\cP}{{\cal P}}
\newcommand{\cG}{{\cal G}}
\newcommand{\cA}{{\cal A}}
\newcommand{\tra}[1]{{\rm tr} ({#1})}

\newcounter{saveeqn}
\newcounter{App} 
\newcommand{\app}{%
\stepcounter{App}%
\setcounter{saveeqn}{\value{equation}}%
\setcounter{equation}{0}%
\renewcommand{\theequation}{\Alph{App}\arabic{equation}} }
\newcommand{\appende}{%
\setcounter{equation}{\value{saveeqn}}%
\renewcommand{\theequation}{\arabic{equation}}  }

\begin{document}
\pagestyle{plain}
\begin{flushright}
{UBCTP 93-13\\LA-UR-93-2991\\
August 1993}
\end{flushright}
\vspace{.4cm}
\renewcommand{\thefootnote}{\alph{footnote}}
\begin{center}
{\Large \bf Consistent axial--like gauge fixing on hypertori}\\
\vspace{2 cm}
{\large Edwin Langmann}\footnote{supported by ``Fonds zur
F\"orderung der wissenschaftlichen Forschung'' under contract Nr.\
J0789-PHY}\\
\vspace{0.3 cm}
{\em Department of Physics, The University of British Columbia\\
Vancouver, B.C., V6T 1Z1, Canada }\\
\vspace{0.8 cm}
{\large Manfred Salmhofer}\footnote{supported by NSERC of Canada}\\
{\em Department of Mathematics, The University of British Columbia\\
Vancouver, B.C., V6T 1Z2, Canada }\\
\vspace{0.8 cm}
{\large Alex Kovner}\\
{\em Theory Division, T-8, MS-B285\\
Los Alamos National Laboratory, Los Alamos, NM 87545}\\

\end{center}
\vspace{0.4cm} \noindent
\setcounter{footnote}{0}

\renewcommand{\thefootnote}{\arabic{footnote}}

\begin{abstract}
We analyze the Gribov problem for $\SU(N)$ and $\U(N)$ Yang-Mills
fields on $d$-dimensional tori, $d=2,3,\ldots$. We give an improved
version of the axial gauge condition and find an infinite, discrete
group $\cG'=\Z^{dr}\rtimes({\Z_2}^{N-1}\rtimes\Z_2)$, where $r=N-1$
for $\GG=\SU(N)$ and $r=N$ for $\GG=\U(N)$, containing all gauge
transformations compatible with that condition. This residual gauge
group $\cG'$ provides (generically) all Gribov copies and allows to
explicitly determine the space of gauge orbits which is an orbifold.
Our results apply to Yang-Mills gauge theories either in the
Lagrangian approach on $d$-dimensional space-time $T^d$, or in the
Hamiltonian approach on $(d+1)$-dimensional space-time $T^d\times \R$.
Using the latter, we argue that our results imply a non-trivial
structure of all physical states in any Yang-Mills theory, especially
if also matter fields are present.

\end{abstract}

\newpage



{}{\bf 1.} The separation of gauge and physical degrees of freedom is
important for understanding gauge theories because the physical
configuration space of a gauge theory is the space of gauge orbits. A
complete gauge fixing should provide a (convenient) choice of
representatives for these orbits. In non-Abelian Yang-Mills theories
this is highly non-trivial. Although in the formal perturbative
expansion the gauge can be fixed by imposing a gauge condition such as
the Coulomb- or the Landau gauge, such a procedure is not really a
gauge fixing because of the existence of Gribov copies, i.e.\ gauge
equivalent field configurations all obeying the gauge condition
\cite{G}. Obviously this is insufficient in the non-perturbative
regime.

Although Gribov ambiguities are absent on $\R^d$ in the axial gauge
\cite{IZ}, they cannot be avoided on $d$-spheres $S^d$ \cite{S,So}, and
the same is to be expected for any other compact manifold $M^d$. As
any deeper investigation of quantum Yang-Mills theory on $\R^d$ will
(in an intermediate step) require the introduction of an infrared
cutoff which can be achieved by replacing $\R^d$ by some compact
manifold $M^d$, the Gribov problem seems closely related to
non-perturbative infrared properties of these models.  Indeed, Gribov
has argued that these ambiguities might give a clue to understanding
confinement \cite{G}.  This argument was much substantiated and
extended by Zwanziger, who analyzed the fundamental domain for the
Landau gauge using a Morse functional \cite{Z1} and showed that it
contains only fields of bounded size. This has important consequences
for the infrared properties of the theory defined such that the fields
are constrained to lie in that region \cite{Z2}.  An explicit
characterization of the fundamental domain for covariant gauges is
difficult because, for topological reasons , there must be some
boundary identifications \cite{B}.

In a recent investigation of gauge theories on a cylinder in the
Hamiltonian framework a complete gauge fixing procedure for $\U(N)$
and $\SU(N)$ Yang-Mills theories on a circle $S^1$ was found
\cite{LS1,LS2}.  Although the Gribov problem prevents a complete gauge
fixing by a linear gauge condition, it was shown that a modified
version of the Coulomb gauge provides a maximal gauge reduction so
that all gauge transformations compatible with that gauge are
(generically) in a {\em discrete} group $\cG'$ independent of the
gauge orbit.  This residual gauge group $\cG'$ allows an explicit
construction of the space of gauge orbits and plays an essential role
for the non-perturbative structure of Yang-Mills theories coupled to
matter on a cylinder \cite{LS2}.

In this paper we analyze the Gribov problem for $\U(N)$ and $\SU(N)$
Yang-Mills gauge theories on $d$-dimensional tori $T^d$, $d \geq 2$.
We give a complete gauge fixing procedure generalizing the one of
\cite{LS1,LS2} to higher dimensions.\footnote{note that $S^1=T^1$, and
that the Coulomb- and the axial gauge in (1+1)-dimensions are
essentially the same} Our interest in the tori is partly motivated by
\cite{MRS}, in which the ultraviolet limit of Yang--Mills theory is
constructed nonperturbatively on a 4-dimensional compact space using
the axial gauge. On the torus the `axial gauge' $A_1=0$ is not really
a gauge condition because it fixes not only gauge but also
physical degrees of freedom, namely the Wilson--Polyakov loops in
$x^1$--direction.  Our gauge fixing procedure uses a consistent
analogue of the axial gauge (consistent in the sense that it really is
a gauge condition). We show by construction that for any Yang-Mills
configuration $A$, there is a gauge transformation $U$ such that
$A^U=U^{-1}(-\ii \, \dd +A)U$ obeys this gauge condition. Moreover,
$U$ is unique up to an element in a residual gauge group $\cG'$, and
for generic field configurations (as explained in detail below), this
group is discrete. It gives all Gribov copies for a given generic
gauge configuration obeying our gauge condition. Moreover, it allows
us to write down the space of gauge orbits explicitly.

Our results apply to Yang-Mills gauge theories either in the
Lagrangian formulation on $d$-dimensional space-time $T^d$, or in the
Hamiltonian approach in $(d+1)$-dimensional space-time $T^d\times \R$.
In the latter case, we give an alternative derivation of our result by
explicitly solving the Gauss' law.  We furthermore argue that the
existence of a non-trivial residual gauge group $\cG'$ implies a
non-trivial structure of all physical states.

{}{\bf 2.} In the following, the structure group $\GG$ of the
Yang-Mills field is $\U(N)$ or $\SU(N)$ in the fundamental
representation and $\gg$ is the Lie algebra of $\GG$. For the
$d$-dimensional torus $T^d$ , $\cA$ is the set of all one--forms
(Yang-Mills field configurations) $A=\sum_{i=1}^d A_i\dd {x^i}$ with
$A_i$ continuous mappings $T^d\to \gg$, and $\cG=Map(T^d;\GG)$ is the
gauge group containig all continuously differentiable mappings
$T^d\to\GG$.\footnote{Our arguments do not change if one imposes
additional, mutually compatible, regularity conditions on the $A_i$'s
and $U$'s.} Parametrizing points $\vec{x} \in T^d$
by $\vec{x}=(x^1,\cdots,x^d)$, $0\leq x^1,\cdots,x^d \leq 2\pi$, we
shall write such functions on $T^d$ as periodic functions on $[0,2\pi
]^d$ with the appropriate continuity and differentiability conditions
at the boundary of $[0,2\pi ]^d$.

We also introduce an algebraic basis in the Lie algebra $\gg$ of
$\GG=\SU(N)$ as follows
\cite{LS2}: Let $e_{ij}$ be the $N\times N$ matrix with the elements
$(e_{ij})_{kl}=\delta_{ik}\delta_{jl}$. We define
$H_i=e_{i,i}-e_{i+1,i+1}$ for $i=1,\ldots, N-1$, spanning the Cartan
subalgebra $\hh$ of $\gg$. Moreover, $E_1^+=e_{1,2}, E_2^+=e_{2,3},
\ldots, E_{N-1}^+=e_{N-1,N}, E_{N}^+=e_{1,3}, \ldots,
E_{\f{1}{2}N(N-1)}^+=e_{1,N}$, and $E_j^-=(E_j^+)^*$ for
$j=1,2,\ldots, \f{1}{2}N(N-1)$.  These matrices obey
\eqa
\label{basis}
\ccr{H_i}{H_j} &=& 0\nonu
\ccr{H_i}{E^{\pm}_j} &=& \pm a_{ij}E^{\pm}_j \qquad\forall i,j
\eqaend
with $a_{ij}$ the elements of $(N-1)\times \f{1}{2}N(N-1)$ matrix
given by $a_{ij}=\delta_{ik(j)}-\delta_{il(j)}-\delta_{i+1,k(j)} +
\delta_{i+1,l(j)}$ where $k(j)$, $l(j)$ are determined from
$E_j^+=e_{k(j),l(j)}$. Moreover, $\ccr{E^+_j}{E^-_k}$ is in $\hh$ if
and only if $j=k$, and
\eq
\label{EE}
\ccr{E^+_k}{E^-_k}=\sum_{j=1}^{N-1}c_{kj}H_j
\eqend
where $c_{kj}$ is a $\f{1}{2}N(N-1)\times (N-1)$-matrix
with 
$c_{kj}=\delta_{kj}$ for $k,j=1,2,\ldots N-1$.

For $\GG=\U(N)$ there is also $H_0=1$. 
Every  $X \in \gg $ has the decomposition
\eq
X=\sum_j X^{0,j}H_j+ \sum_j\left( X^{+,j}E^-_j + X^{-,j}E^+_j\right) ,
\eqend
and $X^{0,j}=(X^{0,j})^*$ and $X^{\pm,j} = (X^{\mp,j})^*$ are uniquely
determined by $X$.

{}{\bf 3.} For simplicity, we start with $d=2$. The generalization to
$d>2$ will be easy.

As mentioned, the axial gauge $A_1=0$ is not really a gauge condition
as it fixes the eigenvalues of the Wilson--Polyakov loops ($\cP$ the
usual path ordering symbol) \eq \label{star}
\cP\Exp{-\ii\int_0^{2\pi}\dd{y^1} A_1(y^1,x^2)} , \eqend which are
gauge--invariant quantities, and in general different from one.  The
simplest similar condition that does not fix them is to require that
$A_1(x^1,x^2)$ is independent of $x^1$ and lies in the Cartan
subalgebra $\hh$ of $\gg$, i.e.\ \eqa \label{1a}
A_1^{0,j}(x^1,x^2)&=&Y_1^j(x^2) \nonu A_1^{\pm,j}(x^1,x^2) &=& 0 \quad
\forall j \: .  \eqaend We prove that this is indeed a consistent
gauge fixing condition by showing that for an arbitrary $A\in\cA$,
there is an $U\in\cG$ such that $A^U$ has the form \Ref{1a}. This
amounts to solving\footnote{$\partial_i\equiv\partial/\partial x^i$}
\eq \label{2}
(A^U)_1(\vec{x})=U^{-1}(\vec{x})[-\ii\partial_1+A_1(\vec{x})]U(\vec{x})
=\sum_j Y_1^j(x^2)H_j\equiv Y_1(x^2) \:, \eqend or equivalently
$[-\ii\partial_1+A_1]U=U Y_1$, which can be solved by the ansatz
$U=u_0 C$ where $u_0$ satisfies the homogeneous eq.\
$[-\ii\partial_1+A_1]u_0=0$ and $C$ obeys $-\ii\partial_1 C= CY_1$.  A
solution of the former for $\vec x \in [0, 2 \pi ]^2$ is provided by
the parallel transporter\footnote{note that this is in general not a
gauge transformation!} \eq \label{ex1} u_0 (x^1,x^2) =
\cP\Exp{-\ii\int_0^{x^1}\dd{y^1} A_1(y^1,x^2)}, \eqend whereas the
general solution of the latter is $C=V\Exp{\ii x^1 Y_1}$ with
$V\in\cG$ independent of $x^1$. Thus \Ref{2} is solved by \eq \label{ex2}
U(x^1,x^2)= u_0(x^1,x^2) V(x^2)\Exp{\ii x^1Y_1(x^2)} \eqend We want
$U$ to be continuously differentiable on the torus, hence $V$ must be
in $Map(T^1;\GG)$. The only other non-trivial condition implying this
is continuity at the endpoints in $x^1$-direction, i.e.\
$U(0,x^2)=U(2\pi,x^2)$ $\forall x^2$. This leads to the following
condition \eq \label{4} V(x^2)^{-1}\cP\Exp{-\ii\int_0^{2\pi}\dd{y^1}
A_1(y^1,x^2)} V(x^2)=\Exp{-\ii2\pi \sum_j Y_1^j(x^2)H_j} \: .  \eqend
Thus $Y_1^j$ are determined by the eigenvalues of the Wilson loops
\Ref{star}, and $V(x^2)\in Map(T^1;\GG)$ has to be such that it
diagonalizes these loops. Obviously such a $V$ exists and \Ref{2} has
always a solution. \Ref{ex2} and \Ref{4} then imply that
$\partial_1 U$ exists and is continuous if $U$ is continuous.

Moreover, the conditions on $A$ and the explicit expressions
\Ref{ex1} and \Ref{ex2} for
$u_0$ and $U$ show that $U$ will be continuously differentiable on the
torus if $V$ and $Y_1$ are continuously differentiable in $x^2$. This
holds if all the diagonal entries of the diagonal matrix on the right
side of \Ref{4} are different.  The gauge condition \Ref{1a} does not
fix the gauge completely. That is to say, \Ref{2} has several
solutions for any given $A_1(x)$.  We now show how to find {\em all}
solutions of \Ref{2}.  We first note that the $Y_1^j(x^2)$ are not
unique but can be shifted by arbitrary integers, \eq \label{5}
Y_1^j(x^2)\to Y_1^j(x^2) + n_1^j,\quad n_1^j\in\Z \: .  \eqend This
can be achieved by the special gauge transformations $\Exp{\ii
x^1\sum_j n_1^jH_j}$.  These transformations provide a representation
of the Abelian group $\Z^{r}$ with $r=N$ for $\GG=\U(N)$ and $r=N-1$
for $\GG=SU(N)$.  Moreover, if $V(x^2)$ diagonalizes the Wilson loops
\Ref{star} then so does $\tilde{V}(x^2)$ if \eq \label{cc1}
\tilde{V}(x^2) = V(x^2)P(x^2) \eqend where $P(x^2)\in
Map(T^1;\PP_{\GG})$ and $\PP_{\GG} =\{ p\in\GG \mid \forall
h\in\hh : p^{-1} h p\in\hh \}$
is the Peter-Weyl group of
$\GG$ (see appendix).  We further note that $P(x^2)\Exp{-\ii
2\pi\sum_j\tilde{Y}_1^j(x^2)H_j }P^{-1}(x^2)=\Exp{-\ii 2\pi\sum_j
Y_1^j(x^2)H_j }$.

In general, \Ref{cc1} is also sufficient, that is, given a particular
solution of \Ref{4} as $V(x^2)$ and $Y_1^j(x^2)$, any other solution
of \Ref{2} has the form $U(x^1,x^2) = U_1(x^1,x^2) P(x^2)$, where \eq
\label{6} U_1(x^1,x^2)= u_0(x^1,x^2) V(x^2) \Exp{\ii
x^1Y_1(x^2)}\Exp{\ii x^1\sum_j n_1^j H_j} \: , \eqend and where
$P(x^2)\in Map(T^1;\PP_{\GG})$ and $n_1^j\in\Z$ $\forall j$ are
arbitrary. If a degeneracy occurs, e.g. if $Y_1^1
(x^2)=Y_1^2(x^2)-Y_1^1(x^2)$ (modulo $\Z$), then $\Exp{-\ii 2\pi\sum_j
Y_1^j(x^2)H_j }$ has two degenerate eigenvalues and therefore commutes
with an $\SU(2)$-subgroup of $\SU(N)$.  More generally, as
$\ccr{E^{\pm}_j}{Y_1(x^1)}=\mp\sum_{i} Y_1^i(x^1) a_{ij}$, the
condition
\eq
\label{c1}
\sum_i Y_1^i(x^1) a_{ij} \in \Z
\eqend
is necessary and sufficient for $\Exp{-\ii 2\pi\sum_j Y_1^j(x^2)H_j}$
to commute with the $\SU(2)$-subgroup generated by $E_j^\pm$, and if
$q$ of these conditions hold there is a $q$-fold degeneracy and an
additional (obvious) $\SU(q+1)$. In the latter case, one has eq.\
\Ref{cc1} with $P(x^2)\in Map(T^1;\PP')$ where $\PP'$ is the group
generated by $\PP_{\GG}$ and this $\SU(q+1)$.

However, the matrices with degenerate eigenvalues form a set of
measure zero in the set of all matrices, and it is conceivable that
this will extend to the functional measure of field configurations
that are degenerate in the above sense, since there is no symmetry
that enforces such a degeneracy. We therefore expect that the
additional gauge transformations associated to these non--generic field
configurations obeying at least one of the conditions \Ref{c1} do not
play a dynamical role (two other reasons will be given in the sequel).

We now show that it is possible (and in fact natural) to further
restrict the gauge freedom by imposing the additional gauge
condition that
\eq
\label{1b}
\int_0^{2\pi}\f{\dd{y^1}}{2\pi} A_2^{0,j}(y^1,x^2) = Y_2^j
\eqend
independent of $x^2$, i.e.\ that the up-to-now arbitrary gauge
transformation $P(x^2)$ can be chosen such that $A^U$ obeys this
condition
(note that this does not restrict the $A_2^{\pm,j}$).
Since $U(\vec{x})= U_1(\vec{x})P(x^2)$
this is equivalent to demonstrating that the following eq.\
has a solution $P(x^2)\in Map(T^1;P)$:
\eq
[-\ii\partial_2 + B_2(x^2)]P(x^2)=P(x^2)\sum_jY^{j}_2 H_j
\eqend
where
\eq
B_2(x^2)\equiv \int_0^{2\pi}\f{\dd{y^1}}{2\pi}\sum_j
(A^{U_1})_2^{0,j}(y^1,x^2)H_j \: .
\eqend
The general solution of this eq.\ is obtained as above
(and path--ordering is not necessary) as $P(x^2)=U_2(x^2) p$
with
\eq
\label{7}
U_2(x^2)=\Exp{-\ii\int_0^{x^2}\dd{y^2}B_2(y^2)}
\Exp{\ii x^2 \sum_j Y^j_2 H_j}\Exp{\ii x^2\sum_j n_2^j H_j}
\eqend
where the $Y_2^j$ are determined by the continuity condition
$P(0)=P(2\pi)$, i.e.\
\eq
\Exp{-\ii\int_0^{2\pi}\dd{y^2}B_2(y^2)} =
\Exp{-\ii 2\pi \sum_j Y^j_2 H_j}
\eqend
and $n_2^j\in\Z$ $\forall j$ and $p\in\PP_{\GG}$ are arbitrary.

The Peter-Weyl group $\PP_{\GG}$ still contains continuous 1-parameter
subgroups (see the appendix). This suggests that it should be possible
to reduce the gauge freedom further, and we do this as follows.
In the appendix we show that for an arbitrary $X\in\gg$, there is a
$p\in\PP_{\GG}$ such that $p^{-1}Xp$ obeys the conditions
\eq
\label{pXp}
(p^{-1}Xp)^{+,j}=(p^{-1}Xp)^{-,j} \quad \forall j=1,2,\ldots N-1 \: .
\eqend
Moreover, for $\GG=\SU(N)$ this $p$ is generically unique up to an
element in the subgroup of $\SU(N)$ generated by the elements $z_i$,
$i=1,2,\cdots N$ and $\sigma$ with matrix elements
\eq
\label{zsigma}
(z_i)_{kl}=\ee{\ii\pi/N}\delta_{kl}\left\{\bma{rc} -1& \mbox{ for $k=i$} \\
1 & \mbox{ otherwise}\ema\right. \qquad (\sigma)_{kl}=\delta_{k,N+1-l}
\eqend
obeying the relations $z_1 z_2\cdots z_N=1$, $\sigma^2=1$,
$z_i^{2N}=1$, $z_i^2=z_j^2$, and $\sigma z_i \sigma = z_{N+1-i}$
$\forall i,j$. This group obviously is a subgroup of
$\Z_{2N}^{N-1}\rtimes \Z_2$ and we denote it as $\ZZ_{\SU(N)}$.  For
$\GG=\U(N)$, $p$ can in addition be multiplied by an arbitrary phase,
hence the corresponding group is $\ZZ_{\U(N)}=\U(1)\times
\ZZ_{\SU(N)}$. In the non-generic case where some of the elements
$X^{\pm,j}$ are zero, there is obviously additional freedom in the
choice of $p$ corresponding to continuous subgroups of $\PP_{\GG}$.

Since $U(\vec{x})=U_1(x^1,x^2) U_2(x^2) p$, we can write
$\int_0^{2\pi}\f{\dd{y^1}}{2\pi}\int_0^{2\pi}\f{\dd{y^2}}{2\pi}
(A^U)_2(y^1,y^2)= p^{-1}X p$ with
$X=\int_0^{2\pi}\f{\dd{y^1}}{2\pi}\int_0^{2\pi}\f{\dd{y^2}}{2\pi}
(A^{U_1U_2})_2(y^1,y^2)$.  It follows that we can impose the
additional gauge condition
\eq
\label{1c}
\int_0^{2\pi}\f{\dd{y^1}}{2\pi}\int_0^{2\pi}\f{\dd{y^2}}{2\pi}
\left(A_2^{+,j}(y^1,y^2)-A_2^{-,j}(y^1,y^2)\right) = 0 \quad \mbox{for
$j=1,2,\ldots N-1$}\: ,
\eqend
and in the generic case this condition fixes the $p\in\PP_{\GG}$ in
eq.\ \Ref{7} up to an element in $\ZZ_{\GG}$.

In the non-generic case where at least one of the following conditions
holds,
\eq
\label{cc2}
\int_0^{2\pi}\f{\dd{y^1}}{2\pi}\int_0^{2\pi}\f{\dd{y^2}}{2\pi}
A_2^{\pm,j}(y^1,y^2) = 0 \mbox{ for $j=1,2,\ldots$ or $N-1$}\: ,
\eqend
there are additional gauge transformations compatible with \Ref{1c}.
By a similar argument as above we expect that these additional gauge
transformations can be ignored.

The group $\ZZ_{\GG}$ obviously contains the center $\CC_{\GG}$ of
$\GG$ ($\Z_{N}$ for $\GG=\SU(N)$ and $U(1)$ for $\GG=U(N)$) which acts
trivially on all Yang-Mills field configurations $A\in\cA$.  For the
Yang-Mills sector the residual gauge freedom associated with the
center is therefore irrelevant and the relevant group of gauge
tranformations compatible with \Ref{1c} is $\ZZ_{\GG}/\CC_{\GG}$
identical with $\Z_{2}^{N-1}\rtimes\Z_2$ for $\GG=\U(N)$ and
$\GG=\SU(N)$. If matter fields are present, invariance under gauge
transformations in the center $\CC_{\GG}$ can be imposed in the matter
sector.

To summarize, we have shown that \Ref{1a}, \Ref{1b} and \Ref{1c}
provide a complete gauge reduction on the torus $T^2$. For every $A\in
\cA$, there is a $U\in\cG$ such that $A^U$ obeys all these conditions,
and for generic $A$ and gauge group $SU(N)$ this $U$ is unique up to
the discrete residual group $\cG'= \Z^{2r}\rtimes ({\Z_2}^{N-1}
\rtimes\Z_2)$. In the {\em generic} case, i.e. for those field
configurations $A$ which do {\em not} satisfy any of the conditions
\Ref{c1} and
\Ref{cc2}, $\cG'$ provides all Gribov copies of our gauge condition.
For non-generic field configurations obeying at least one of these
conditions one has additional Gribov copies, and it is straightforward
to write down a complete list of these.  However, as argued already
above, we do not consider these additional Gribov copies as important.
This is also suggested by the fact that the latter are coordinate
dependent in the sense that they change if we replace the $x^1$- by
the $x^2$-directions, i.e.  diagonalize the Wilson loops in
2-direction instead of those in 1-direction, or impose the condition
\Ref{1c} on other $j's$, hence they do not have a gauge invariant
meaning. This suggests that the additional Gribov copies associated
with these field configurations can be regarded as arising from
`coordinate singularities', in contrast to those corresponding to the
residual gauge group $\cG'$ which do have a gauge invariant
significance.

{}{\bf 4.} For $d>2$, the complete gauge fixing condition is given by
\eqa
\label{2a}
A_1^{0,j}(x^1,x^2,\cdots,x^d)&=& Y_1^j(x^2,\cdots,x^d)\nonu
A_1^{\pm,j}(x^1,x^2,\cdots,x^d)&=& 0\quad \forall j
\eqaend
\eqa
\label{2b}
\int_0^{2\pi}\f{\dd{y^1}}{2\pi}
A_2^{0,j}(y^1,x^2,\cdots,x^d) &=& Y_2^j(x^3,\cdots,x^d) \nonu
\int_0^{2\pi}\f{\dd{y^1}}{2\pi}\int_0^{2\pi}\f{\dd{y^2}}{2\pi}
A_3^{0,j}(y^1,y^2,x^3,\cdots,x^d) &=& Y_3^j(x^4,\cdots,x^d) \nonu
&\vdots&\nonu
\int_0^{2\pi}\f{\dd{y^1}}{2\pi}\int_0^{2\pi}\f{\dd{y^2}}{2\pi}
\cdots\int_0^{2\pi}\f{\dd{y^{d-1}}}{2\pi}
A_d^{0,j}(y^1,y^2,,\cdots,y^{d-1},x^d) &=& Y_d^j\quad \forall j
\eqaend
\eqa
\label{2c}
\int_0^{2\pi}\f{\dd{y^1}}{2\pi}\int_0^{2\pi}\f{\dd{y^2}}{2\pi}
\cdots\int_0^{2\pi}\f{\dd{y^d}}{2\pi}
\left(A_d^{+,j}(y^1,\cdots,y^d) -A_d^{-,j}(y^1,\cdots,y^d)\right)=0
\nonu
\mbox{for $j=1,2,\ldots,N-1$} \: .
\eqaend
It is straightforward to extend our argument for $d=2$ above and
construct for an arbitrary $A\in\cA$ the $U\in\cG$ such that $A^U$
obeys these conditions, and to show that this $U$ is unique up to an
element in the residual gauge group
$\cG'=\Z^{dr}\rtimes ({\Z_2}^{N-1} \rtimes \Z_2)$.

It is easy to write down explicitly the action of all elements in
$\cG'$ on Yang-Mills configurations $A$ obeying
\Ref{2a}--\Ref{2c}. As the resulting list of equations is quite
lengthy and not very illuminating, we refrain from doing this here.

The result for $d=1$ is somewhat different \cite{LS2} but easily
recoverd as follows: in this case there is only one gauge condition
$A_1(x^1)=\sum_j Y_1^j H_j$, and the space of all configurations
obeying this condition is obviously $\R^r$. Moreover, the gauge
transformation $U$ transforming a general $A\in\cA$ to one obeying
this condition is generically unique up to an element in
$\Z^r\rtimes\PP_{\GG}$, but as $\HH_{\GG}\subset\PP_{\GG}$ acts
trivially on all field configuration obeying the gauge condition, the
residual gauge group is $\cG'=\Z^r\rtimes\PP_{\GG}/\HH_{\GG}
=\Z^r\rtimes S_N$ (see appendix).

{}{\bf 5.} Our result above allows us to determine the space of all
gauge orbits explicitly: Let $\cA_0$ be the set of all $A\in\cA$
obeying the gauge condition
\Ref{2a}--\Ref{2c}. Then obviously the orbit space $\cA/\cG$ is
(generically) identical to $\cA_0/\cG'=(\cA_0/\Z^{dr})/({\Z_2}^{N-1}
\rtimes \Z_2 )$. $\cA_0$ is a manifold, but as there are field
configurations in $\cA_0$ that are fixed points of $\sigma$
\Ref{zsigma}, $\cA_0/\cG'$ is only an orbifold, and thus has
singularities which cannot be removed by a mere choice of coordinates.
Thus some of the elements in the residual gauge group, or in other
words some of the Gribov copies, produce more than just coordinate
singularities. This agrees with the results of Babelon and Viallet
\cite{BV}.

{}{\bf 6.} In the following, we discuss Yang-Mills theory on
space-time $T^d \times \R$ in the Hamiltonian framework. In this case,
the consistency of the gauge fixing conditions can also be seen from
the Gauss' law $G=\sum_{i=1}^d(-\ii\partial_i E_i +
\ccr{A_i}{E_i})+\rho=0$ (with $\rho$ the temporal component of some
matter current as usual).

To avoid clumsy notation, the following discussion is given for $d=2$;
the extension to $d>2$ is again trivial.  Introducing the notation
\neqa
X(n_1,x^2) = \int_0^{2\pi}\f{\dd{y^1}}{2\pi}\ee{-\ii n_1y^1}X(y^1,x^2) \\
X(n_1,n_2) = \int_0^{2\pi}\f{\dd{y^2}}{2\pi}\ee{-\ii n_2y^2}X(n_1,y^2)
\neqaend
and using the basis $\{H_j,E^{\pm,j} \}$ in $\gg$ introduced above,
the gauge condition \Ref{1a} allows to write the Gauss' law as
\eqa
\ii n_1E_1^{0,j}(n_1,x^2) + \partial_2E_2^{0,j}(n_1,x^2)+
\tilde\rho^{0,j}(n_1,x^2) = 0\nonu
\ii\left(n_1\pm \sum_i Y_1^i(x^2)a_{ij}\right)E_1^{\pm,j}(n_1,x^2) +
\partial_2E_2^{\pm,j}(n_1,x^2)+ \tilde\rho^{\pm,j}(n_1,x^2) = 0
\eqaend
with
\eq
\tilde\rho = \rho + \ccr{A_2}{E_2} \: .
\eqend
These equations can be solved (in the generic case),
\eqa
\label{3a}
E_1^{0,j}(n_1,x^2)= -\f{\tilde\rho^{0,j}(n_1,x^2)-\ii\partial_2
E_2^{0,j}(n_1,x^2)}{\ii n_1}
\quad \forall n_1\neq 0\nonu
E_1^{\pm,j}(n_1,x^2) = -\f{\tilde\rho^{\pm,j}(n_1,x^2)-\ii\partial_2
E_2^{\pm,j}(n_1,x^2)} {\ii\left(n_1\pm \sum_i Y_1^i(x^2)
a_{ij}\right)}\quad \forall n_1\in\Z \: ,
\eqaend
and determine those and only those components of $E_1$ conjugate to
components of $A_1$ set to zero by the gauge condition \Ref{1a}. This
shows that \Ref{1a} is a consistent gauge condition.

By now we have taken into account all components of the Gauss' law
except $G^{0,j}(n_1=0,x^2)=0$. Using the gauge condition \Ref{1b}, we
can write the latter as
\eq
\ii n_2E_2^{0,j}(n_1=0,n_2) + \tilde\rho^{0,j}(n_1=0,n_2) = 0 \: ,
\eqend
which can be solved
\eq
\label{3b}
E_2^{0,j}(n_1=0,n_2) = -\f{\tilde\rho^{0,j}(n_1=0,n_2)}{\ii
n_2}\quad\forall n_2\neq 0 \: ,
\eqend
thus showing that also \Ref{1b} is consistent.  We are finally left
with the Gauss' law component $G^{0,j}(n_1=0,n_2=0)$.
Using \Ref{EE}, this can be written as
\eqa
\sum_{k=1}^{\f{1}{2}N(N-1)}c_{kj}\left[E_2^{+,k}A_2^{-,k} -
E_2^{-,k}A_2^{+,k}\right](n_1=0,n_2=0)
+ \rho^{0,j}(n_1=0,n_2=0)\nonu
\equiv  \left( E_2^{+,j}(0,0)A_2^{-,j}(0,0) -
E_2^{-,j}(0,0)A_2^{+,j}(0,0) \right) + \{\cdots\} = 0
\eqaend
(we used $c_{kj}=\delta_{kj}$ for $k,j=1,2,\ldots ,N-1$) where
$\{\cdots\}$ does not depend on $E(A)_2^{\pm,j}(0,0)\equiv
E(A)_2^{\pm,j}(n_1=0,n_2=0)$. With the gauge condition \Ref{1c}, i.e.\
$A_2^{+,j}(0,0)=A_2^{-,j}(0,0)$, this is (in the generic case) easily
solved for the component of $E_2$ conjugate to
$A_2^{+,j}(0,0)-A_2^{-,j}(0,0)$,
\eq
\label{3c}
E_2^{+,j}(0,0)-E_2^{-,j}(0,0)= -\f{\{\cdots\}}{A_2^{+,j}(0,0)}
\eqend
showing that also \Ref{1c} is consistent.

One can insert \Ref{3a}, \Ref{3b} and
\Ref{3c} into the Hamiltonian for Yang-Mills theory coupled to matter.
Similarly as discussed in \cite{LS2} for $d=1$, one thereby obtains
the Hamiltonian in the restricted Hilbert space $\cH'$ obtained by
restricting the Yang-Mills configurations to $\cA_0$, i.e.\ imposing
\Ref{1a}, \Ref{1b} and
\Ref{1c}.  On $\cH'$ we still are left with a representation of the
residual gauge group $\cG'$. As the latter is a discrete group and has
a convenient semi-direct product structure, one can explicitly
construct all states in $\cH'$ which are invariant under $\cG'$, and
it is this set of states which span the physical Hilbert space
$\cH_{\rm phys}$ of the model
\cite{LS2}.

It is interesing to note that in the present approach the non-generic
field configurations obeying at least one of the conditions \Ref{c1}
and \Ref{cc2} appear as singularities in the eqs.\ \Ref{3a}, \Ref{3c}.
As shown in detail in Ref.\ \cite{LS1} for $d=1$ (see eq.\ (18)), they
lead to singularities in the matter ($\rho$--$\rho$) interactions on
the physical Hilbert space $\cH_{\rm phys}$ of the model. As these
singuar interactions are repulsive, we believe that they are not
`dangerous' but an intrinsic feature of non-Abelian Yang-Mills
theories.\footnote{we thank E.Seiler for discussions about this}
 This is supported by the fact that the effective
Hamiltonian for Yang-Mills theory on a cylinder coupled to an {\em
external} matter current $\rho(x)=\rho\delta(x^1)$ reduces to a
completely integrable Sutherland Hamiltonian\footnote{E.L. is grateful
to A. Gorsky for a discussion on this point} \cite{s}
\eq
H=-\f{1}{8\pi}
\sum_{j,k}b_{jk}\f{\partial^2}{\partial Y_1^j \partial Y_1^k} +
\f{\pi}{2}\sum_j\left(
\f{{\rho}^{+,j}{\rho}^{-,j}}
{\sin^2(\pi\mbox{$\sum_i$}Y_1^ia_{ij})^2} +
\f{{\rho}^{-,j}{\rho}^{+,j}}
{\sin^2(\pi\mbox{$\sum_i$}Y_1^ia_{ij})^2}
 \right)
\eqend
(we set the coupling constant to 1) where $b_{kl}=\tra{H_j H_k}$ \cite{LS1}
(insert into eq.\ (18) of Ref.\ \cite{LS1} and use the formula
$\sum_{n\in\Z}1/(n-y)^2=\pi^2/\sin^2(\pi y)$; to see for
$\GG=\U(N)$ that this is indeed a Sutherland Hamiltonian
\cite{s} use the basis $(H_i)_{kl}=\delta_{ik}\delta_{il}$.
For $\GG=\SU(N)$ one can also use this basis leading to the same
result with the additional constraint $\sum_{j=1}^{N}Y_1^j=0$ which
just fixes the `center-of-mass motion' of the $Y_1$'s).
This Hamiltonian is positive, and it is self-adjoint
in spite of the singularities \cite{s}.

{}{\bf 7.} We have given a complete gauge fixing procedure for $\U(N)$
and $\SU(N)$ gauge theories on tori $T^d$. It is similar to the axial
gauge but does not fix physical degrees of freedom. By complete we
mean that the residual gauge group is generically discrete. It would
be interesting to find a similar procedure for other manifolds $M^d$,
e.g.\ $M^d=S^d$. In Ref.\ \cite{So} it is proven that on spheres there
is no gauge condition with a residual gauge group which is discrete
and the same for {\em all} gauge field configurations obeying the
condition. We note that our result would not contradict a similar
theorem on tori due to the existence of the non-generic field
configurations with larger residual gauge groups containing continuous
subgroups.

One hopes that in the limit in which all linear dimensions of a
manifold become infinite, one obtains the same `thermodynamic limit'
for any reasonable manifold. We note that the proper treatment of the
axial gauge is even non-trivial on an infinite space $\R^d$ \cite{C}.
In our opinion, even in that case the axial gauge cannot be fixed
completely, the reason being basically the same as for the case we
have discussed.

The gauge transformation $U$ that maps every $A$ to its representative
obeying our gauge condition is a product of transformations $U_1 \cdot
\ldots \cdot U_d$ applied consecutively.  The map $U_1$ diagonalizes
the Wilson--loops in $x^1$--direction and is similar in spirit to a
rewriting of the theory in terms of loop variables \cite{L}, but we do
not try to give a complete reformulation of the theory in terms of the
latter. This partial rewriting seems advantageous because the Jacobian
of the complete change of variables has not been calculated, whereas
in our case it can be calculated as the product of $d$ Jacobians.

For {\em pure} Yang-Mills theory not only peridic functions
$[2,\pi]^d\to\GG$ correspond to gauge transformations, but in fact
also all those which are periodic only up to an element in the center
$\CC_{\GG}$. However, these latter transformations must not be fixed
as they correspond to physical symmetries of the system which may be
broken, and our construction applies also to this case.



\app
\begin{center}
\section*{APPENDIX}
\end{center}
In this appendix we (a) explicitly construct the
Peter-Weyl group
\eq
\PP_{\GG}=\{ p\in\GG \mid \forall h\in\hh : p^{-1} h p\in\hh  \}
\eqend
and (b) prove its properties needed for the final gauge reduction
in the eqs.\ following \Ref{1c}.

{\bf (a)} As $\U(N)=\U(1)\times\SU(N)$ implies
$\PP_{\U(N)}=\U(1)\times \PP_{\SU(N)}$ it suffices to consider
$\GG=\SU(N)$. Then $\hh$ consists of all traceless diagonal real matrices
$h={\rm diag}\{ h_1,\ldots ,h_N\}$ (and the $h_i$ are the eigenvalues of $h$).
Let $p\in\PP$, and take any $h \in \hh$ such that all $h_i$ are different.
Since $\tilde h = php^{-1}$ and $h$ have the same eigenvalues,
there is a permutation $\pi$ of $(1,2,\ldots,N)$ such that
$\tilde h = \hat \pi h \hat \pi ^{-1}$
whith $\hat \pi \in SU(N)$ representing $\pi$,
$(\hat\pi)_{ij} = e^{\ii \pi / N deg(\pi )} \delta_{i,\pi (j)} $
and $deg (\pi) =0 $ for $\pi $ even and $1$ otherwise. So
$H=\hat \pi ^{-1} p $ commutes with $h$ and thus is diagonal.
Since $p$ and $\hat \pi $ are in $SU(N)$,
$\det H = 1 $ and so $H\in \HH$, and $p$ has the
decomposition $p=\hat \pi H$.
As obviously
$\hat\pi^{-1}H\hat\pi\in\HH$ $\forall H\in\HH$, $\pi\in S_N$,
and since the representation $\pi \mapsto \hat\pi $ is faithful,
\eq
\PP_{\SU(N)}=\HH_{\SU(N)} \rtimes S_N\: .
\eqend

{\bf (b)} By definition of $E_i^{\pm}$, \Ref{pXp} is equivalent to
\eq
\label{b1}
(p^{-1}Xp)_{i,i+1}=(p^{-1}Xp)_{i+1,i} \mbox{ for $i=1,2,\ldots N-1$}
\eqend
 for $X\in\gg$ and $p\in\PP$. $X$ is hermitean and $p$ unitary, so
\Ref{b1} is a reality condition on either side. Since by (a) $\PP$
is a semidirect product, every $p \in \PP$ has a unique decomposition
$p=H \hat \sigma$ with $H \in \HH$ and $\sigma \in S_N$, and
we can solve \Ref{b1} for $H$ and $\sigma $ separately.
$X \in \gg $ is arbitrary, so we can assume $X_{i,i+1} \neq 0$
for all $i$, thus $X_{k,k+1} = R_k e^{\ii \xi_k}$ with $R_k > 0$ and
$\xi_k \in \R$. Inserting $H_{ij} = \delta_{ij} e^{\ii \alpha_i}$
into \Ref{b1} gives
\eq
\Exp{2\ii(\xi_i-\alpha_i+\alpha_{i+1})}=1 \mbox{ for $i=1,2,\ldots N-1$} \: .
\eqend
which amounts to the recursion
$\alpha_{k+1} = \alpha_{k} - \xi_k + m_k \pi $ with
$m_k \in \{ 0,1\}$. Denoting
\eq
\label{b2}
\alpha^{(0)}_1=\f{1}{N}\sum_{i=1}^{N-1}(N-i)\xi_i,
\quad \alpha^{(0)}_i = \alpha_1 -
\sum_{l=1}^{i-1} \xi_l\quad\mbox{ for $i\in \{2,3,\ldots N\}$}\: ,
\eqend
all $p\in\HH$ obeying \Ref{b1} are given by
$p=H_0 z \zeta$ with
\eq
\label{b3}
(H_0)_{ij}=\delta_{ij}\Exp{\ii\alpha_i^{(0)}}
\eqend
and
\eq
\label{b4}
z_{ij}=\delta_{ij}\lambda\ee{\ii\pi\ell_i} \mbox{ with $\ell_1=0$,
$\ell_i\in\{0,1\}$ for $i \geq 2$ and
$\lambda = \Exp{-\ii\pi/N\sum_{k=2}^N\ell_k}$,}
 \eqend
and $\zeta \in \Z_N$ for $G=SU(N)$, $\zeta \in U(1) $ for $G=U(N)$.
$\sigma\in S_N$ is determined by
$(\hat\sigma^{-1}Z\hat\sigma)_{i,i+1}=(\hat\sigma^{-1}Z\hat\sigma)_{i+1,i}$ for
$Z_{i,i+1}=Z_{i+1,i}$.
Taking $Z_{i,j+1} \neq Z_{j+1,i}$ for any $(i,j) $ with $i \neq j$,
gives the condition $\sigma (i+1) = \sigma (i) \pm 1 $,
the only two solutions of which are
$\sigma (i)= i $ and $\sigma (i) = N+1-i$.
We conclude that all $p\in\PP$
obeying \Ref{b1} are given by $p=H_0 z \zeta$ where $z$ is in the subgroup
of $\PP$ generated by the $z$ eq. \Ref{b4}, $\hat\sigma$
and $\zeta $ is in the center of the group. We denote
this latter group as $\ZZ_{\SU(N)}$.

\appende

\begin{center}\subsubsection*{Acknowledgement}\end{center} We are
grateful to V. Rivasseau, E. Seiler, and G. Semenoff for helpful
discussions and comments. E.L. acknowledges the hospitality of the
theory division T-8 of Los Alamos National Laboratories where part of
this work was done.

\end{document}